\documentstyle[]{article}
\newcommand{\beq}{\begin{equation}}
\newcommand{\eeq}{\end{equation}}
\newcommand{\tr}{{\rm tr}}
\newcommand{\dcl}{D^{cl}}

\newcommand{\cle}{{\cal L}^{eff}}
\newcommand{\spc}{\;\;\; , \;\;\; }
\newcommand{\calv}{{\cal V}}
\def\anp#1#2#3{Annals Phys. #1 (#3) #2}
\def\ibid#1#2#3{{\it ibid.} #1 (#3) #2}
\def\npb#1#2#3{Nucl. Phys. B #1 (#3) #2}
\def\npsb#1#2#3{Nucl. Phys. Proc. Suppl. B #1 (#3) #2}
\def\plb#1#2#3{Phys. Lett. B #1 (#3) #2}
\def\prc#1#2#3{Phys. Rev. C #1 (#3) #2}
\def\prd#1#2#3{Phys. Rev. D #1 (#3) #2}
\def\prl#1#2#3{Phys. Rev. Lett. #1 (#3) #2}
\def\phr#1#2#3{Phys. Rep. #1 (#3) #2}
\def\zpc#1#2#3{Z. Phys. C #1 (#3) #2}
\begin{document}
\begin{center}
{\Large \bf Notes on the Deconfining Phase Transition}
\vskip2.pc
{\large \bf Robert D. Pisarski}
\vskip.5pc
{\it High Energy Theory\\
Department of Physics\\Brookhaven National Laboratory\\
Upton, NY 11973 USA\\
pisarski@quark.phy.bnl.gov}
\end{center}
\vskip.5pc
\begin{abstract}
I review the deconfining phase transition in an $SU(N)$ gauge theory
without quarks.  After computing the interface
tension between $Z(N)$ degenerate vacua deep in the deconfined phase,
I follow Giovannangeli and Korthals Altes, and suggest a new
model for (discrete) Polyakov loop spins.
Effective theories for (continuous) Polyakov loop spins are constructed,
including  those with $Z(N)$ charge greater than one,
and compared with Lattice data.  
About the deconfining transition, the expectation
values of $Z(N)$ singlet fields
(``quarkless baryons'') may change markedly.  Speculations include:
a possible duality between Polyakov loop and ordinary spins
in four dimensions, and how $Z(N)$ bubbles might be
guaranteed to have positive pressure.  
\end{abstract}
\vskip1.pc
\begin{center}
{\it To appear in the Proceedings of
``QCD Perspectives on Hot and Dense Matter'', a
NATO Advanced Study Institute, August, 2001, Cargese, France}
\end{center}
\section{Overview}

In these lectures I review the deconfining
phase transition in a ``pure'' $SU(N)$ gauge theory, without dynamical
quarks.  
Gauge theories are ubiquitous in physics, so
their phase transitions are manifestly of fundamental importance.
Two examples may include
the collisions of large nuclei at high energy and the early universe.

The phase transitions of gauge theories without quarks
are of especial interest,  since then the order parameter,
and many other aspects of the phase transition, 
can be characterized precisely \cite{polyakov,hooft,svetitsky}.  
While of course QCD includes quarks, this is not
an academic exercise.  
Recent results from the Lattice
on ``flavor independence'' --- for both the pressure \cite{flavorind,karsch}
and quark susceptibilities \cite{gavai_sucep} ---
suggest that the results from the pure glue theory may be, 
in a surprising and unexpected fashion, relevant for QCD.
(Whether flavor independence can be generalized when
there are many light flavors is not known.)

Albeit indirectly, the Lattice has already told us much
about what happens in a pure gauge theory with three colors.
By asymptotic freedom, at infinite temperature the
pressure is that for an ideal gas of gluons.
In the confined phase, the pressure is very small, essentially zero.
So the question is, as the pressure turns on at the transition
temperature $T_c$, how rapidly does it approach the ideal gas limit?
The Lattice tells us relatively quickly: by about $2 T_c$,
it is already $80\%$ of the way to ideality.  
The ``$2$'' in $2T_c$ is meant schematically; it is certainly
not, say, $10 T_c$.  Above $2T_c$, the pressure then approaches
ideality slowly, from below.

\pagestyle{myheadings}
\markright{Notes on Deconfinement}

This suggests that from temperatures of $2 T_c$ on up, that
the theory is some sort of quasiparticle gas of thermal quarks and gluons;
{\it i.e.}, a Quark-Gluon Plasma.
By this I mean that after suitable dressing from bare into quasi-particles,
the residual interactions are weak.  
This is seen from the Lattice:
like the ratio of the true to the ideal pressure,
the ratio of (gauge-invariant) masses to the temperature
also vary slowly above $2 T_c$ \cite{laineph,gavai_masses}.

It is known that for the free energy, 
direct perturbative calculations fail at astronomically
high temperatures \cite{free}; there is only a perturbative
Quark-Gluon Plasma above temperatures of $\sim 10^7$~GeV. 
Thus from temperatures of $10^7$~GeV down to $2 T_c$, the
theory is what I call a non-perturbative Quark-Gluon Plasma. 

One approach to the non-perturbative QGP is to 
fold in the effects of Debye screening.
It is known that the pressure, 
as obtained from the Lattice, can be fit to an ideal
gas of massive gluons all the way down to $T_c$ \cite{quasi}.
The problem is that as introduced, these masses aren't gauge
invariant.  Further, 
in the end one is just
fitting one function of temperature, the pressure, to another,
thermal masses.
(Still, it is most intriguing that these fit thermal masses become large 
near $T_c$.)

Another approach is provided by the resummation of
Hard Thermal Loops (HTL) \cite{htl,rebhan,blaizot,braaten}.
For the gauge field, if $A_0$ is the time-like component
of the vector potential, the Debye mass term is just
$\sim \tr(A_0^2)$.  For gluons,
HTL's are the gauge invariant, analytic continuation of this mass term
from imaginary to real time.  At present, though, HTL resummation
has been used mainly to compute the pressure.  The crucial test,
yet remaining, are the results which it gives for
Polyakov loop correlation functions \cite{laineph,yaffe}.  

A method to compute all static correlation functions in the
non-perturbative Quark-Gluon Plasma starts with the
(perturbative)
construction of an effective theory in three dimensions \cite{kajantie}.  
From the original gauge theory at $T\neq 0$,
static magnetic fields produce three dimensional gluons,
while static electric fields give $A_0$, as an adjoint
scalar coupled to these gluons.
Due to the power-like infrared divergences of gauge theories in three
dimensions, perturbative calculations are useless in this effective
theory.  Since the effective theory is purely bosonic, though,
static correlation functions can be efficiently computed
by numerical means on the Lattice.
While heroic perturbative calculations are required, 
for three colors this method appears to work
down to $2 T_c$, where the method itself  indicates its failure.
To be fair, this method only yields static correlation
functions, and not those in real time.  With HTL resummation, this
continuation is almost automatic.

What about below $2 T_c$?  If the transition is strongly first
order, then presumably the quasiparticle regime extends all of
the way down to $T_c$.  Lattice data suggests that for four
colors \cite{four}, the transition is strongly first order.  
(Although data on correlation lengths near $T_c$ is absent.)  For
more than four colors, if the deconfining transition is like the
Potts model, then it becomes more strongly first order as $N$ increases.
On the Lattice, ``reduced'' models have a first order transition
at infinite $N$ \cite{largeN}.  In the continuum, for really no
good reason, I remain unconvinced \cite{second}.

For two colors, however, the deconfining
transition is almost certainly of second
order \cite{two}.  For three colors, the transition is of first
order \cite{three}, as predicted by Svetitsky and
Yaffe \cite{svetitsky}.
Recent results suggest, however, that the transition is so
weakly first order that it 
is more accurate to speak of a ``nearly second order'' transition
\cite{bielefeld_mass}.

This is seen most clearly not from the pressure, but
from the behavior of electric and magnetic masses.  I define
the electric, $m_{el}$,  and magnetic masses,
$m_{mag}$, in a gauge invariant way, from the fall-off of the
two-point functions for Polyakov loops, and the (trace of the)
magnetic field squared, respectively.  
In the perturbative QGP, $m_{el} > m_{mag}$ \cite{laineph}, while
in the non-perturbative QGP, $m_{el} \sim m_{mag}$ 
\cite{laineph,gavai_masses}.  In contrast, 
in the transition regime $m_{mag}/T$ is approximately
constant, but $m_{el}/T$ appears to 
drop by a factor of ten as $T: 2T_c \rightarrow T_c^+$ \cite{bielefeld_mass}.
A similar drop in the string tension is seen as $T: 0 \rightarrow T_c^-$
\cite{bielefeld_mass}.

The broad outlines of the appropriate effective theory in
the transition region for a (nearly)
second order deconfining transition have been known for some time
\cite{polyakov,hooft,svetitsky,banks,patel}.  In the QGP 
regime, one deals with $A_0$.  In the transition
region, one trades $A_0$ in for the thermal Wilson line.  
Traces of powers of the thermal Wilson line
give Polyakov loops, which are gauge invariant.

The effective theory of Polyakov loops is just beginning
\cite{loop,loop1,loop2,loop3,loop4,loop5,loop6}. 
If true, instead of the $A_0$ quasiparticles of the
QGP, near $T_c$ it is more useful to view the theory
as a {\it condensate} of Polyakov loops.
This could produce dramatic signatures in heavy ion collisions,
with hadronization at $T_c$ computed semiclassically from
the decay of Polyakov loop condensates \cite{loop1,loop2,loop3}.

In these Lectures I  provide some background to
understand these questions.  After
explaining why $SU(N)$ gauge theories have $Z(N)$ degenerate
vacua \cite{hooft}, 
I compute the interface tension between these vacua
at high temperature
\cite{interface,altes,interface_other,partition,altes_talk,gka,notnot,not}.
Viewing the $Z(N)$ vacua as discrete spins, and
using results of Giovannangeli and Korthals-Altes
\cite{gka}, I propose a new spin model, distinct
from the usual Potts model.  (The order of the transition seems to agree
with Potts for $N \leq 4$, but is unknown for $N>4$.)
I then develop an effective theory of Polyakov loop spins,
considered as continuous variables.  Even with the limited amount of
relevant Lattice data which exists at present,
the form of this effective theory is sharply constrained.
Especially intriguing is the possibility that
the expectation values of fields which are singlets
under the $Z(N)$ symmetry --- which I term
quarkless baryons --- change suddenly about $T_c$.

The lectures are in part pedagogical, in part base speculation.
The latter includes a possible
duality in four dimensions, and how $Z(N)$ bubbles,
which in perturbation theory 
can have negative pressure \cite{not},
might be ensured of positive pressure  non-perturbatively.
I also review what is known about renormalization of Wilson loops
and Wilson lines \cite{renorm_cont,gava,enq_90,bielefeld_ren}.  

In these lectures I only discuss the Polyakov Loops Model
\cite{loop,loop1,loop2,loop3,loop4,loop5,loop6} in passing.
I hope to provide a basis for understanding its motivation.  At present,
a major unsolved problem is the analytic continuation
from imaginary to real time.
In abelian gauge theories, the analytic continuation of
the Debye mass term gives the Random Phase Approximation \cite{tsvelik};
in nonabelian theories, it gives Hard Thermal Loops \cite{htl}.
Absent any results whatsoever, I do have the temerity
to coin a phrase for the analytic continuation of the Polyakov
Loops Model to real time, 
as the ``{\it Nonabelian Random Phase Approximation}''.
Time (dependence) will tell.

\subsection{Saturation}

My motivation for studying this subject is its relevance for 
the collisions of large nuclei at very high energies.  Thus
at the outset, I wish to add some general comments about nucleus-nucleus
collisions, and especially how it might relate to models of saturation.

Consider a completely implausible situation, the collision of two neutron
stars (say) at relativistic energies.  
For a neutron star, the transverse area is essentially infinite
on nuclear scales.  If the stars completely overlap (zero impact
parameter), then at very high energies, a nearly baryon-free region
is generated between them (about zero rapidity).  
The system starts out with energy density, but no
pressure.  It is then reasonable to think that the system builds up pressure,
and thermalizes, before it flies apart.  

The crucial question for the collisions of two large nuclei is whether
for gold or lead nuclei, with $A \sim 200$, that
this finite value of $A$ is close to infinity, or 
represents some intermediate regime.
This will be decided by experiment, 
at the SPS, RHIC, and the LHC.  After one year of running, it
appears clear that {\it something} dramatic has happened between SPS
and RHIC energies \cite{exp}.  Precisely what is still being sorted out.

I wish to comment here on the relevance of
``saturation'' at these energies \cite{saturation}. 
At a very pedestrian level,
this can be viewed as a type of finite size effect at $A < \infty$.
Consider a collision in the rest frame of one nuclei.
The diameter of the other nucleus, with $A \sim 200$, is $\sim 15$~fm.
This distance becomes Lorentz contracted.  Thus we can
ask, in the rest frame of one nuclei, when does the incident
nuclei look like a pancake of negligible width?  
We want the distance to be really small on typical hadronic scales.
If a typical hadronic scale is $1$~fm, then, a small scale might be
$1/4 \rightarrow 1/3$~fm, say.  To contract $15$~fm  down to
these sizes then requires a center of mass energy per nucleon pair,
$\sqrt{s}/A$, on the order of $\sim (3\rightarrow 4) \times 15 =
45\rightarrow60$~GeV.  While an extremely naive estimate, this
does seem to be a reasonable estimate for the
$\sqrt{s}/A$ where the distribution of particles in $AA$ collisions changes 
dramatically, developing a ``central plateau''
as a function of pseudo-rapidity.  Thus perhaps the appearance
of the central plateau is where the effects of saturation first appear.

The details are far more involved, but 
but our understanding of saturation, 
which is known formally as the Color Glass \cite{saturation},
gives us some confidence that the system
is described in terms of ``saturated'' gluons, with a
characteristic momentum $p_{sat}\sim 1-2$~GeV.  

The Color Glass changes all assumptions in one fundamental
respect.  Following Bjorken, the usual assumption
is that the system thermalizes on
a ``typical'' hadronic time scale, $\sim 1$~fm/c.  If saturation kicks
in, however, all typical scales are then given in terms of $1/p_{sat}$,
which is a {\it much} smaller time scale, $\sim .1-.2$~fm/c.  
If this is the relevant scale, then evolution to a thermal state,
for a nucleus of size $6$~fm, appears much more plausible.
Certainly it yields testable predictions, which
are testable experimentally.

\section{$Z(N)$ symmetries in $SU(N)$}

I start by reviewing how, following
't Hooft \cite{hooft}, a global $Z(N)$ symmetry emerges from
a local $SU(N)$ gauge theory.  The action, including quarks, is
\beq
{\cal L} = \frac{1}{2} \; {\rm tr} \; G^2_{\mu \nu} \; + \;
\overline{q} \; i \not \!\! D \, q \; ,
\eeq
where
\beq
D_\mu = \partial_\mu - i g A_\mu \;\;\; , \;\;\;
G_{\mu \nu} \; = \; \frac{1}{- i g} \; [D_\mu,D_\nu] \; ;
\eeq
$A_\mu = A^a_\mu t^a$, with the generators of $SU(N)$ normalized
as ${\rm tr}(t^a t^b) = \delta^{ab}/2$.  The Lagrangian
is invariant under $SU(N)$ gauge transformations $\Omega$,
\beq
D_\mu \rightarrow \Omega^\dagger D_\mu \Omega \;\;\; , \;\;\;
q \rightarrow \Omega^\dagger q \; .
\eeq
As an element of $SU(N)$, $\Omega$ satisfies
\beq
\Omega^\dagger \Omega = {\bf 1} \; , \;
\det \Omega = 1 \; .
\eeq
Here $\Omega$, as a local gauge transformation, is a function of
space-time.  

There is one especially simple gauge transformation ---  a constant
phase times the unit matrix:
\beq
\Omega_c \; = \; e^{i \phi} \; {\bf 1} \; .
\eeq
To be an element of $SU(N)$, the determinant must be one, which requires
\beq
\phi = \frac{2 \pi j}{N} \;\;\; , \;\;\;
j = 0, 1\ldots (N-1) \; .
\label{phi}
\eeq
Since an integer cannot change continuously from point to point,
this defines a global $Z(N)$ symmetry.  

\subsection{$Z(N)$ at nonzero temperature}

As a particular
gauge transformation, $Z(N)$ rotations are always
a symmetry of the Lagrangian, either with or without quarks.
I now show that with quarks, they are not a symmetry of the theory,
because they violate the requisite boundary conditions.

I work in Euclidean space-time at a temperature $T$, so the
imaginary time coordinate $\tau$, is of finite
extent, $\tau: 0 \rightarrow \beta = 1/T$.  
The proper boundary conditions in imaginary time are dictated
by the quantum statistics which the fields must satisfy.  As
bosons, gluons must be periodic in $\tau$; as fermions,
quarks must be anti-periodic:
\beq
A_\mu(\vec{x},\beta) \; = \; + A_\mu(\vec{x},0) \;\;\; , \;\;\;
q(\vec{x},\beta) \; = \; - q(\vec{x},0) \; .
\eeq
Obviously any gauge transformation which is periodic in $\tau$
respects these boundary conditions.  't Hooft noticed, however,
that one can consider more general gauge transformations which
are only periodic up to $\Omega_c$:
\beq
\Omega(\vec{x},\beta) = \Omega_c \;\;\; , \;\;\;
\Omega(\vec{x},0) = 1 \; .
\label{e21}
\eeq
Color adjoint fields are invariant under this transformation,
while those in the fundamental representation are not:
\beq
A^\Omega(\vec{x},\beta) \; = \; 
\Omega_c^\dagger A_\mu(\vec{x},\beta) \Omega_c \; = \;
A_\mu(\vec{x},\beta) \; = \; + \; A_\mu(\vec{x},0) \; ,
\eeq
\beq
q^\Omega(\vec{x},\beta) \; = \; \Omega_c^\dagger \, q(\vec{x},\beta)\; 
= \; e^{- i \phi} q(\vec{x},\beta) \;
\neq - \; q(\vec{x},0) \; .
\eeq
Here I have used the fact that $\Omega_c$, as a constant phase
times the unit matrix, commutes with any $SU(N)$ matrix.
Consequently, pure $SU(N)$ gauge theories have a global $Z(N)$
symmetry which is spoiled by the addition of dynamical quarks.

In the pure glue theory, an order parameter for
the $Z(N)$ symmetry is constructed using the thermal Wilson line:
\beq
{\bf L}(\vec{x}) = {\bf P} \exp 
\left( i g \int^\beta_0 A_0(\vec{x},\tau) d \tau \right) \; ;
\eeq
$g$ is the gauge coupling constant, and $A_0$ the vector
potential in the time direction.  The symbol $\bf P$ denotes
path ordering, so that the thermal Wilson line transforms
like an adjoint field under local $SU(N)$ gauge transformations:
\beq
{\bf L}(\vec{x}) \rightarrow \Omega^\dagger(\vec{x},\beta) \;
{\bf L}(\vec{x}) \; \Omega^\dagger(\vec{x},0) \; .
\eeq
The Polyakov loop \cite{polyakov} is the trace of the thermal
Wilson line, and is then gauge invariant:
\beq
\ell = \frac{1}{N} \; {\rm tr} \, {\bf L} \; .
\eeq
Under a global $Z(N)$ transformation, the Polyakov loop $\ell_1$
transforms as a field with charge one:
\beq
\ell \rightarrow e^{i \phi} \ell \; .
\label{e222}
\eeq

At very high temperature, the theory
is nearly ideal, so $g\approx 0$, and naively one expects that
$\langle \ell \rangle \sim 1$.  
Instead, the allowed vacua exhibit a $N$-fold degeneracy:
\beq
\langle \ell \rangle \; = \;
\exp \left( \frac{2 \pi i j}{N} \right) \; \ell_0  \;\;\; , \;\;\;
j \; = \; 0,1\ldots(N-1) \; ,
\eeq
defining $\ell_0$ to be real; $\ell_0 \rightarrow 1$
as $T \rightarrow \infty$.  Any value of $j$ is equally good,
and signals the spontaneous breakdown of the global $Z(N)$ symmetry.

At zero temperature,
confinement implies that $\ell_0$ vanishes \cite{hooft}.
The modulus, $\ell_0$, is nonzero above $T_c$:
\beq
\ell_0 \; = \; 0 \;\;\; , \;\;\; T<T_c
\;\;\; ; \;\;\; \ell_0 \; > \; 0 \;\;\; , \;\;\; T>T_c \; .
\eeq
As is standard, 
if $\ell_0$ turns on continuously at $T_c$,
the transition is of second order; if it jumps at $T_c$, it
is of first order.  What is atypical is that the $Z(N)$ symmetry
is broken at high, instead of low, temperatures.  For a heuristic
explanation of this in terms of $Z(N)$ spins, see sec. (3.2).

One can also understand what it means to say that the global
$Z(N)$ symmetry is violated by the presence of dynamical quarks.
In the pure glue theory, $Z(N)$ rotations take us from one
degenerate vacua to another, all of which have the same
pressure; see sec. (3.2).
Adding dynamical quarks (with real masses),
the stable vacuum is that for which $\langle \ell \rangle$
is real, $j=0$ \cite{banks,patel}.
As discussed in sec. (3.4), because of quarks the pressure for a $Z(N)$
state with $j \neq 0$ is less than the stable vacuum; 
thermodynamically, they are unstable.
What is exciting to some of us \cite{notnot},
however, is the possibility that these 
$Z(N)$ rotated states might be {\it meta}stable.  Such
``$Z(N)$ bubbles'' could have cosmologically interesting 
consequences \cite{notnot}.
Others contest whether any of this makes any sense \cite{not}. 

The usual interpretation of the Polyakov
loop is as the free energy of an infinitely heavy test quark \cite{mcsvet}:
\beq
\langle \ell \rangle = \exp\left(- F_{test}/T\right) \; .
\label{e12}
\eeq
This cannot be quite right: when $N=2$, the left hand side 
can be of either sign, while for $N \geq 3$, it is complex.
In contrast, free energies are real, so the right hand side is positive.

I suggest a different view.
Consider the propagator for a scalar in a background gauge
field (the extension to fermions is irrelevant here).  This
propagator is given by a Feynman sum over paths:
\beq
\Delta = \int {\cal D}x^\mu
\exp \left( - \int ds \left(
\frac{\dot{x}^2}{2} \; + \; m \; + \; i g A^\mu \dot{x}^\mu \right)\right)\; ,
\eeq
with $s$ the length of the path for the worldline of the particle,
and $\dot{x} = dx^\mu/ds$.
A very heavy quark moves in a straight line; 
in imaginary time, it sits wherever you put it.
As a colored field, however, it also 
carries a color Aharonov-Bohm phase.  
This phase is nontrivial, and is precisely the thermal Wilson line.
Thus: {\it the Polyakov loop, $\ell$, 
is the trace of the propagator for a test quark}. 

Confinement then means that (the trace of) this propagator vanishes.
For two colors, for example, the confining vacuum is $Z(2)$
symmetric: it is composed of domains, of definite size, in which
$\ell = +1$ and $\ell = -1$.  As the test quark travels through
each domain, it picks up one phase or the other.  Over a very
long path, these phases cancel out, giving zero overall.  The
same holds for higher $N$, except that there are then $N$ types
of domains.  This picture appears analogous to the localization of
an electron in a random potential \cite{anderson}.

$Z(N)$ rotations can be expressed in terms of the canonical
formalism \cite{hooft,partition}.
In $A_0 = 0$ gauge, usually the partition function is strictly
a trace of $\exp(-{\cal H}/T)$ ($\cal H$ is the corresponding Hamiltonian), 
sandwiched between the same state:
\beq
Z = \Sigma \; \langle \psi | \exp(-{\cal H}/T) | \psi \rangle .
\eeq
The sum is over all gauge invariant states in the theory.  In a canonical
formalism, one inserts a projector to ensure that the states are gauge
invariant, satisfying Gauss' Law, although the process is
standard.  Instead, what enters here is a "twisted" trace:
\beq
Z(\Omega_c) = \Sigma \; \langle \psi^{\Omega_c} | 
\exp(-{\cal H}/T) | \psi \rangle .
\eeq
Here, $\psi^{\Omega_c}$ represents the gauge transform of $\psi$ by the
gauge transformation $\Omega_c$.  This is not automatically equal to
$\psi$, because Gauss' Law only ensures that the state is invariant under
local gauge transformations, and does not restrict its behavior under the
global gauge transformations.
This twisted trace is only possible in a gauge theory.

\section{$Z(N)$ Vacua and Bubbles}

\subsection{Tunneling between Degenerate $Z(N)$ Vacua}

Typically, discussions of gauge theories at nonzero temperature
work up from zero temperature.  But the
zero temperature theory confines, which is complicated.  Instead,
I work down from infinite temperature, in a gas of nearly ideal gluons.
I now compute the amplitude to tunnel from one $Z(N)$ vacua
to another by semi-classical means
\cite{interface,altes,interface_other,gka}. 

For simplicity, I work with two colors.  To compute the interface
tension, I put the system in a box:
\beq
\tau: 0 \rightarrow \beta \;\;\; , \;\;\;
x,y: 0 \rightarrow L_t \;\;\; , \;\;\;
z: 0 \rightarrow L \; .
\eeq
I choose one arbitary direction, say the z-direction, and make that
much longer than the transverse spatial directions, $x$ and $y$,
and than the direction in imaginary time, $\tau$.  I impose
boundary conditions such that the Polyakov loop 
has one value in $Z(2)$ at one end of the box, and the other
value at the other end of the box:
\beq
\ell(0) = 1 \;\;\; , \;\;\; \ell(L) = -1 \; .
\eeq
With these boundary conditions, the system is forced to form
an interface between the two ends of the box.   The simplest interface
is one which is constant in the transverse directions.  Thus
the natural expectation is that with the above boundary conditions,
the action is
\beq
S_{inter} \; = \; L^2_t \frac{c}{g^2}\;\;\; ,\;\;\;
c \; = \; c_0 \; + \; g c_1 \; + \ldots
\eeq 
One expects the result to start as $\sim 1/g^2$, as a semiclassical
probability for tunneling in weak coupling.  This is standard with
instantons, {\it etc.}  Higher order corrections to the leading
term, $c_0$, are generated by including quantum effects, and
produce the corrections $c_1$, {\it etc.}

As an ansatze for the $Z(N)$ interface, I take
\beq
A_0^{cl}(z) \; = \; \frac{\pi T}{g}\;  q(z) \; \sigma_3 \;\;\; , \;\;\;
\sigma_3 \; = \;
\left( 
\begin{array}{cc}
1 & 0  \\
0 & -1 
\end{array} 
\right)
\label{e22a}
\eeq
With this ansatz, the Polyakov loop is
\beq
\ell(z) \; = \; \cos \left( \pi q(z) \right) \; .
\label{e22b}
\eeq
Thus the boundary conditions are satisfied by taking
\beq
q(0) \, = \, 0 \;\;\; , \;\;\; q(L) \, = \, 1 \; .
\eeq

At the classical level, the action of the above configuration is:
$$
{\cal S}^{cl} \; = \; \int^\beta_0 d\tau \;
\int d^3x \; \frac{1}{2} \; {\rm tr}\;
\left( \left(G_{\mu \nu}^{cl}\right)^2 \right)
$$
\beq
 \;\;\;\;\;\; \;\;\;\;\;\; \;\;\;\;\;\; \;\;\;\;\;\; \;\;\;\;\;\; 
 = \; L^2_t \; \frac{2 \pi^2 T}{g^2} \; 
\int \; dz \left( \frac{dq}{dz} \right)^2 \; .
\label{e22}
\eeq
Unsurprisingly, the action for the gauge field becomes
a kinetic term for the classical field.  There is only
a kinetic term, since the classical field commutes with
itself.  

This implies, however, that at the classical level, there is
{\it no} difference between the two vacua, or indeed any
state with $q \neq 0$!  One can take $q(z)= z/L$; then
the action is $\sim L/L^2 \sim 1/L$, and vanishes as
$L \rightarrow \infty$.  In terms of the above,
\beq
c_0 \; = \; 0 \; .
\eeq

I now show that quantum corrections generate a nonzero value
for $c_1$.  Since $c_0$ vanishes, this is then the leading
term in a semiclassical expansion; the tunneling probability
is then not $\sim 1/g^2$, but only $\sim 1/g$.

To show this, it is necessary to compute the quantum corrections
about the above semiclassical configuration, taking
\beq
A_\mu \; = \; A_\mu^{cl} \; + \; A_\mu^q \; .
\eeq
The computation is a bit involved, but 
an excellent exercise in the use of the background field
method \cite{background}, which is always good to know.

It is convenient to take background field gauge,
\beq
D^{cl}_\mu A_\mu^q \; = \; \partial_\mu A_\mu^q \; - \; i g 
[A^{cl}_\mu,A^q_\mu] \; = \; 0 \; .
\eeq
I work in euclidean space-time with $(++++)$ metric.  Note
that the appropriate covariant derivative is that in
the adjoint representation.  With this gauge fixing,
the Lagriangian density is
\beq
{\cal L} = \frac{1}{2} \; {\rm tr} 
\left( G^{cl}\right)^2 
\; + \; \frac{1}{\xi} \; \tr \left( D^{cl} \cdot A^q\right)^2
+ \overline{\eta} \left( - D^{cl}\cdot D \right) \eta \; ,
\eeq
suppressing ugly vector indices.

With this form, it is easy integrating out the quantum fields to
one loop order, and obtain the quantum action
\beq
{\cal S}^q(A^{cl}) \; = \;
\frac{1}{2} \; \tr \log \left( \Delta^{-1}_{\mu \nu} \right)
\; - \; \tr \log \left( \Delta^{-1}_\eta \right) \; .
\label{e23}
\eeq
At one loop order, the full effective action is the
sum of the classical action, (\ref{e22}), and the quantum
action, (\ref{e23}).  The quantum action involves 
the inverse propagators in a background field.  That for
the gluon is
\beq
\Delta^{-1}_{\mu \nu} = 
\; - D_{cl}^2 \delta_{\mu \nu} \; + \; (1 - \xi^{-1} ) D^{cl}_\mu
D^{cl}_\nu \; + \; 2 i g \; [G^{cl}_{\mu \nu},\;\;] \; ,
\eeq
while that for the ghost is (to lowest order in $g$)
\beq
\Delta^{-1}_\eta \; = \; - D_{cl}^2  \; .
\eeq
These results are valid for an arbitrary background gauge field.

I now make a crucial assumption, and assume that the field
$q(z)$ is constant in space.  For the relevant tunneling amplitude,
in fact $q(z)$ does depend upon $z$; what happens, however,
is that for the quantum action, this variation only enters to
higher order in the coupling constant.  

This assumption vastly simplifies the problem.  Since $A_0$
lies in the $\sigma_3$ direction, it 
is a diagonal matrix, and covariant derivatives commute:
\beq
[\dcl_\mu , \dcl_\nu] \; \sim \; G^{cl}_{\mu \nu} \; = \; 0 \; .
\eeq
For example, one can easily show that the quantum action is
independent of the gauge fixing condition.  The variation of the quantum action
with respect to the gauge fixing parameter $\xi$ is
\beq
\frac{\partial}{\partial \xi^{-1}} \; {\cal S}^q
\; = \; \frac{1}{2} \; \tr \left( - \dcl_\mu \dcl_\nu
\Delta^{cl}_{\mu \nu} \right) \; ,
\eeq
with $\Delta^{cl}_{\mu \nu}$ the gluon propagator.
Normally, this is difficult to compute.  In the present example,
however, if covariant derivatives can be assumed to commute with each
other, then they can be treated just like ordinary derivates, so that
\beq
\Delta^{cl}_{\mu \nu} \; = \; \frac{\delta^{\mu \nu}}{-D_{cl}^2}
\; + \; (1 - \xi) \frac{D_{cl}^\mu D_{cl}^\nu}{(-D_{cl}^2)^2} \; .
\eeq
Consequently,
\beq
\frac{\partial}{\partial \xi^{-1}} \; {\cal S}^q
\; = \; \frac{\xi}{2} \; \tr (1) \; .
\eeq
Thus there is gauge dependence in the quantum action, but it is completely
independent of the background field, and so can be safely ignored.

Consequently, I take background Feynman gauge,  $\xi = 1$, and
\beq
{\cal S}^q \; = \; \tr \; \log \left(- D_{cl}^2 \right) \;.
\eeq
The overall factor is expected for a massless gauge field, with
two (spin) degrees of freedom.

To compute the determinant in this background field, I introduce
the ``ladder'' basis,
\beq
\sigma^+ \; = \; \frac{1}{\sqrt{2}} \; \left(
\begin{array}{cc}
0 & 1  \\
0 & 0 \\
\end{array}
\right) \;\;\; , \;\;\; 
\sigma^- \; = \; \frac{1}{\sqrt{2}} \; \left(
\begin{array}{cc}
0 & 0 \\
1 & 0 \\
\end{array}
\right) \; .
\eeq
This is useful because of the commutation relations:
\beq
[\sigma_3 , \sigma^\pm] \; = \; \pm \; 2 \sigma^\pm \; ,
\eeq
so that the covariant derivative becomes
\beq
D^{cl}_0 \sigma^\mp \; = \;
\left( \partial_0 - i g \left( \frac{\pi T}{g} q \right) [\sigma_3, \; ] 
\right) \sigma^\mp \; = \;
i \; (2 \pi T) \; ( n \pm q ) \sigma^\mp \; .
\eeq
In the last expression, I have gone to momentum space.  Remember that
given the periodic boundary conditions at nonzero temperature,
\beq
k_0 \; = \; i \; 2 \pi T \; n\; ,
\eeq 
where for a bosonic field, such as a gluon, the periodic boundary
conditions require that $n$ be an integer, $n= 0, \pm 1, \pm 2 \ldots$.
In the present case, it is handy to introduce the shifted momentum,
\beq
k_0^\pm \; = \; i \; 2 \pi T \; (n\pm q)\; .
\eeq 
In the trace, the sign of $q$ doesn't matter, so that
in momentum space,
\beq
{\cal S}^q \; = \; 2 \; \tr \; \log \left( (k_0^+)^2 + \vec{k}^2 \right) \; .
\eeq

In computing integrals at nonzero temperature, the usual
approach is to do the sum over the $k_0$'s first by contour
integration or the like, and then
integrate over the spatial momentum.
In the present example, instead it is better to first
integrate over the spatial momentum, and then sum over the
$k_0$'s, using zeta functions tricks.

Only the variation of the action with respect to $q$, 
\beq
\frac{\partial}{\partial q} {\cal S}^q \; = \;
8 \pi T \; (\beta L_t^2 L) \; T \sum_{n =-\infty}^{+ \infty}
\; \int \frac{d^3k}{(2\pi)^3} \; \frac{k_0^+}{(k_0^+)^2 + \vec{k}^2} \; ,
\label{e13}
\eeq
is needed.
The integral over $\vec{k}$ can be done using dimensional regularization,
viewing $k_0^+$ like a mass.
Using the standard integral,
\beq
\int d^nk \; \frac{1}{(k^2 + m^2)^a} \; = \;
\frac{\Gamma(a - n/2)}{\Gamma(a)}
\; \frac{\pi^{n/2} }{(m^2)^{a - n/2}} \; ,
\eeq
the result is finite:
$$
\frac{\partial}{\partial q} {\cal S}^q \; = \;
8 \pi L_t^2 L \; T \sum_{n =-\infty}^{+ \infty} \;
k_0^+ \left( \frac{-1}{4 \pi} | k_0^+| \right) \; 
$$
\beq
\;\;\;\;\;\;\;\;\;\;\;\;\;\;\;\;\;\;\;\;\;\;\;\;\;\;
= \; - 8 \pi^2 T^3 L_t^2 L \; \sum_{n = -\infty}^{+ \infty} \;
(n+q) |n+q| \; .
\eeq
The sum over $n$, where
$n$ runs from minus infinity to plus infinity,  is turned
into a sum from zero to plus infinity:
\beq
= - 8 \pi^2 T^3 L_t^2 L \; \sum_{n = 0}^{+ \infty} \;
\left( (n+q)^2 - (n + (1-q))^2 \right) \;
\eeq
While these sums are very divergent, mathematicians
know how to handle them.  They are defined by the analytic
continuation of the Riemann zeta-function:
\beq
\zeta(z,q) \; = \; \sum_{n=0}^{+\infty} \; \frac{1}{(n+q)^z} \; .
\eeq
Using
\beq
\zeta(-2,q) \; = \; - \; 
\frac{1}{12} \; \frac{d}{dq} \left( q^2 (1-q)^2 \right) \; ,
\eeq
gives
\beq
{\cal S}^q \; = \;
L^2_t \; \frac{4 \pi^2 T^3}{3} \; 
\int dz \; q^2 \; (1-q)^2 \; .
\eeq
This expression is only valid for $q:0 \rightarrow 1$; given the
derivation, it is a periodic function in $q$ with
period one.  Also note that in anticipation of later results,
I have replaced a factor of the length $L$ by $\int dz$.

Physically, the computation of quantum corrections has lifted the
degeneracy in $q$ by generating a potential for $q$.  
In the full effective action, it helps to rescale the length
in the $z$ direction, introducing
\beq
z' \; = \; \sqrt{\frac{2}{3}} \; g T \; z \; .
\eeq
With this rescaling, the complete effective action at  one-loop order is
\beq
{\cal S}^{eff} \; = \;
{\cal S}^{cl} \; + \; {\cal S}^{q} \; = \;
L_t^2 \; \; \frac{4 \pi^2}{\sqrt{6}} \frac{T^2}{g} \;
{\cal S}' \; ,
\eeq
where
\beq
{\cal S}' \; = \; \int \; dz' \;
\left( \left( \frac{dq}{dz'} \right)^2 \; + \;
q^2 \, (1-q)^2 \right) \; .
\eeq
The factor of 
$T^2$ follows on dimensional grounds, as at high temperature,
$T$ is the only natural mass scale in the problem.  (The
renormalization mass scale doesn't appear until 
next to leading order \cite{interface,altes}.)  

What is most interesting is that the $1/g^2$, which
we had expected, becomes only a $1/g$.  This is
because our effective action only acquires a potential at one
loop order.

This begs the important question, why should this
effective action be trusted to one loop order?  What about
the effects to higher loop order?  The point is that in the
new effective action, the relevant distance scale is not
just $1/T$, but $1/(gT)$: notice the factor of $gT$ in the
definition of the rescaled length $z'$.  Thus for small $g$,
any variations in the effective action occur over much
larger distance scales than $1/T$.  This is why our method
of derivation --- ignoring the variation of $q(z)$ in the
quantum action ----  works.  It does vary in
space, but in weak coupling, this variation is very slow,
and can be ignored.

Having reduced the effective action to the above form, we
merely want the ``kink'' which interpolates between the
two vacua.  While the general form of the kink is well known,
in fact we only need the action.  To compute the action, we
note that the ``energy'' $\epsilon$ for this system is
conserved:
\beq
\epsilon \; = \; \left( \frac{dq}{dz'} \right)^2 - q^2 (1-q)^2 \; .
\eeq
Here we view the spatial coordinate $z'$ as a kind of time;
saying the energy is conserved means that it is independent of $z'$.
The energy is then a constant of motion; with these boundary conditions,
this energy vanishes.  Zero energy implies
\beq
\frac{dq}{dz'} \; = \; q(1-q) \; .
\eeq
Using this, 
\beq
{\cal S}' \; = \; 
2 \; \int \; dz' \; q^2 (1-q)^2 \; = \;
2 \; \int^1_0 \; dq \; q (1-q) \; = \; \frac{1}{3} \; .
\eeq
Putting everything together,
\beq
{\cal S}^{eff} \; = \;
L_t^2 \; \frac{4 \pi^2}{3 \sqrt{6}} \; \frac{T^2}{g} \; ,
\eeq
which is the final result for two colors.  

This demonstrates that $c_1 \neq 0$; the interface tension
vanishes classically, but is generated through quantum effects.
The most interesting feature of the analysis is how
the tunneling amplitude goes from the expected $1/g^2$ to
just $1/g$.  There are examples in 
string theory where tunneling amplitudes are not $1/g^2$, but
only $1/g$.  These examples appear special to string theory,
as the appearance of the coupling constant in this fashion is
more or less natural.  That is not the case here; the $1/g$ is
really novel.  The appearance of the inverse distance scale $gT$ is
reasonable, as the Debye screening mass in a thermal bath. 

\subsection{Spins and Polyakov Loop Spins: Duality?}

Remember the behavior of a usual Ising magnet, in which 
spins $\sigma_i = \pm$ interact through a
coupling constant $J_{mag}$.  The Hamiltonian is:
\beq
{\cal H} \; = \; - \; J_{mag} \; 
\Sigma_{i,\hat{n}} \; \sigma_i \cdot\sigma_{i+\hat{n}}
\; .
\eeq
The sum is over all lattice sites, $i$,
and nearest neighbors to $i$, $\hat{n}$. 
The spins align at low temperatures, and disorder at high 
temperatures.  This is just because
the partition function is ${\cal Z} \sim \exp(-{\cal H}/T)$.
In magnets, the spin-spin
coupling is more or less independent of temperature, so that
with $\sim J_{mag}/T$ in the exponential, ordering
wins at low temperature, and loses at high temperature.  

The interface tension can easily be estimated.  
The simplest interface is to take all spins on the left hand
side spin up, and all on the right hand side, spin down.
If $a$ is the lattice spacing, then the interface tension,
defined as above, is $\sim J$; by construction, its width is
$a$.  This very sharp interface is not the
configuration of lowest energy, 
but the true interface tension is of order $\sim J$, with
a width of order, $a$.  Another example, more familiar
to field theorists, is given by a scalar field with 
a double well potential \cite{altes_talk}.

Now consider an effective lagrangian
for Polyakov loops.  Over distances $> 1/T$, 
the four-dimensional theory reduces to an effective theory
of spins in three spatial dimensions.  
For two colors, Polyakov loops are a type of $Z(2)$ spin,
with $\ell = \pm$.  From
the above ${\cal S}^{eff}$, 
\beq
J_{Polyakov} \; \sim \; \frac{T^2}{g} \; .
\eeq

Now it is easy to understand why Polyakov loop spins order
at high, instead of low, temperature.
For magnets, the partition function involves $\exp(-J_{mag}/T)$;
for Polyakov loop spins, instead we have
$\exp(- J_{Polyakov}/T)$; but as 
$J_{Polyakov} \sim T^2$, in all the temperature dependence
in the exponential is not $1/T$, as for ordinary magnets,
but $T$!

This also leads to a natural conjecture of  duality: that the temperature
for Polyakov loop spins, and an ordinary magnet, are related as
\beq
T_{Polyakov} \; \sim \; \frac{1}{T_{mag}} \; .
\eeq
I have assumed that the variation of the gauge coupling constant
with temperature can be neglected.  

This argument is extremely heuristic, and carries an important
qualification.  For ordinary spins, the lattice spacing is of
course fixed.  (This is true as well for a scalar field with
double well potential \cite{altes_talk}.)  
From the derivation of the interface tension above, however,
the width of the interface is the inverse Debye mass, $\sim 1/(gT)$.
Thus the argument fails in the limit of high temperature,
since then the size of any single domain is becoming very large,
$\sim \sqrt{\log(T)}/T$, as $T\rightarrow \infty$.  This doesn't contradict
the conclusion of ordering at high temperature, since any single
domain is, by definition, an ordered state.

Now assume that 
the transition is of second order, as happens for two colors \cite{two}.
Then the correlation length diverges at $T_c$; with Polyakov loop spins,
it decreases as $T$ increases from $T_c$.  This
divergence is determined as usual by scaling at a critical point.
Even so, the
underlying length scale which fixes the lattice spacing for
effective Polyakov loop spins is fixed, set by a mass scale
proportional to $T_c$, {\it etc.}
Near $T_c$, we have implicitly made the assumption
that the interface tension remains proportional to $\sim T^2$.
This cannot be true very near $T_c$, since for a second order
transition, the interface tension must vanish at $T_c$.

So is the argument of any use?  Well, if $d$ is the number
of space-time dimensions, then simply on geometric grounds,
$J_{Polyakov} \sim T^{d-2}$.
In three dimensions, then, $J_{Polyakov} \sim T$;
depending upon the value of $J_{Polyakov}/T$, the system can still order
above $T_c$, so there is no obvious contradiction.  However,
it does suggest that the {\it width} of the critical region
is much {\it narrower} in four, as opposed to three, dimensions.
For $SU(2)$ gauge theories, this comparison 
is of interest in its own right.

For a strongly first order transition, at first
sight one might think that one
should be able to directly check if 
$J_{Polyakov} \sim T^{d-2}$.  This is complicated by the fact that
near $T_c$, $Z(N)$ states don't tunnel directly from one to another,
but from one $Z(N)$ state, to the symmetric
vacuum, to another $Z(N)$ state.  

\subsection{Polyakov Loop Spins: Potts versus GKA}

The above analysis can be extended to more than two colors.
For two colors, 
there is only one interface tension, between $\ell = +1$ and
$\ell = -1$.  For $N$ colors, 
the vacuum is one of the $N$th roots of unity, 
$\ell = \exp(2 \pi i j/N)$, $j = 1\ldots(N-1)$.  By charge
conjugation, under which $\ell \rightarrow \ell^*$, 
the states $j$ and $N-j$ are equivalent.  There are then about
$\sim N/2$ distinct interface tensions.

At any $N$, the smallest interface tension is between  $j=0$
and $j=1$.  Defining ${\cal S}^{eff} = \alpha_1 L_t^2$,
then at next to leading order,
\beq
\alpha_1 
\; = \; \frac{4 (N-1) \pi^2}{3 \sqrt{3N}} \,
\frac{T^2 }{g(T)}
\, \left( 1 - 12.9954... \; \frac{g^2 \, N}{(4 \pi)^2}  + \ldots \right)  \,,
\eeq
where the running coupling constant $g^2(T)$ is defined using
a modified $\overline{MS}$ scheme \cite{gka}.

I remark that Boorstein and Kutasov \cite{interface_other} 
argued that due to infrared divergences,
$\sigma_1$ is not $\sim 1/g$, but one
over the magnetic mass scale, $\sigma_1 \sim 1/g^2$.  While
hardly conclusive, at least at next to leading order,
there are no sign of infrared divergences.

The interface tension from $j=0$ to $j=k$
has been computed by
Giovannangeli and Korthals Altes (GKA) \cite{gka}.  
The result is amazingly simple:
\beq
\alpha_k \; = \; \frac{k \, (N - k)}{N-1} \; \alpha_1 \; .
\eeq

Now I construct an effective theory of discrete $Z(N)$ spins.
I forget about the factors of temperature which preoccupied me
in the previous subsection; all that I am concerned with
is the dependence on the distance between the $Z(N)$ spins.
This suggests what I term the GKA model.  The spins
at each site of the lattice are integers $j$, $j=0\ldots(N-1)$;
the Hamiltonian is
\beq
{\cal H}_{GKA} \; = \;  J_{GKA} \; \Sigma_{j_i } \; k \; (N-k)
\; \;\; , \;\;\; k = |j_i - j_{i + \hat{n}}|_{ mod \; N} \; .
\eeq
Tracing through the factors of $N$, and holding $g^2 N$ fixed
as $N \rightarrow \infty$, the coupling constant $J_{GKA}>0$
is of order one as $N\rightarrow \infty$.

The GKA model is in contrast to the Potts model, with Hamiltonian 
\beq
{\cal H}_{Potts} \; =  \; J \; \Sigma_{j_i} \; \delta_{k0}
\; .
\eeq
For the Potts model with $J>0$, the energy is lowered if
if two spins are equal, while if they differ ---
no matter by how much --- the energy vanishes.

For two and three colors, there is no difference between the GKA
model and the Potts model.  For example, consider the case of three
colors.  Then $j=1$ is equivalent to $j=2$, so there is only
one interface.  That is, for the 
three roots of unity, any root is right next to the other two.

The Potts model is known to be of first
order for any number of states greater than, or equal to, three.
For the GKA model, 
in mean field theory the transition is of
first order for four colors \cite{gka}; after
all, the interaction between $j=0$ and $j=2$ is $4/3$ that
between $j=0$ and $j=1$.  Thus it would be very surprising
if the GKA model wasn't also of first order when $N=4$.
Further, Lattice simulations of $SU(4)$ find a first order
deconfining transition \cite{four}.

As the number of colors increases, though, the Potts and GKA
models become increasingly different.
Whatever $N$ is, in the Potts model any spin state interacts equally
strongly with any other spin state.  In the GKA model, at large $N$
spins only interact significantly with those which are close
in spin space.  For example, $\sigma_1 \sim N$, while $\sigma_{j} \sim
N^2$ for $j\sim N/2$.  It would be interesting to know the
order of the phase transition in the GKA model at large $N$,
both in mean field theory and numerically.

This assumes that the interaction between Polyakov loop
spins --- computed in the limit of very high temperature ---
remains the same all of the way down to $T_c$.  This certainly
is wrong for a second order transition, but the question here
is if it is first order.  Thus: does the
interface tension stay $\sim j(N-j)$ (higher powers of
$\sim j^2(N-j)^2$ only make the more less like Potts), or become constant,
independent of $j$? 

\subsection{$Z(N)$ Bubbles}

From the one-loop effective action, we can define a ``potential''
for $q$ due to gluons, ${\cal V}_{gl}$.  I will be schematic,
suppressing all inessential details, and taking two colors
for now.  From the computation of
the one-loop effective action, it is clear that the assumption
of $L_t \ll L$ was actually a matter of words; one obtains
identically the same result in an infinite volume.  Thus from ${\cal S}^q$,
I define
\beq
{\cal V}_{gl}(q) \; \sim \; T^4 \; \left( q^2 (1-q)^2 \; + \;
f_g \right) \; ;
\eeq
$f_g T^4$ is proportional to
the free energy of gluons at a temperature $T$.
As noted above, this potential is only valid in the region $0\leq q \leq 1$;
it is periodic, with period one, outside of this region:
\beq
{\cal V}_{gl}(q+1) \; = \; {\cal V}_{gl}(q) \; .
\eeq
Obviously, $q=0$ and $q=1$ are degenerate,
\beq
{\cal V}_{gl}(0) \; = \; {\cal V}_{gl}(1) \; .
\eeq
This follows from the $Z(2)$ symmetry of the pure glue theory.

The quark contribution is computed similarly:
\beq
{\cal V}_{qk}(q) \; \sim \; T^4 \; \left( 2 \, q^2  \; - \; q^4 \; + \;
f_{qk} \right) \; ;
\eeq
$f_qk T^4$ is proportional to the quark contribution to the free energy.
This expression is only valid for $0 \geq q \geq 1$; else $q$
is defined modulo one.  Consequently, 
\beq
{\cal V}_{qk}(q+2) \; = \; {\cal V}_{qk}(q) \; .
\eeq
This must be true for any potential, since $q=0$ and
$q=2$ both give $\ell = +1$.  Moreover, while $q=1$ is a 
an extremal point of the potential, it is a local maximum,
not a minimum, with 
\beq
{\cal V}_{qk}(1) \; > \; {\cal V}_{qk}(0) \; .
\eeq
This shows how quarks 
violate the global $Z(2)$ symmetry of the two color theory.  

To one loop order,
the total potential for $q$ is the sum of the gluon and
quark contributions.  With many (light) quark flavors, the total
potential is like the quark contribution, with a maximum at $q=1$.  
Dixit and Ogilvie \cite{notnot} first noticed that if
the number of quark flavors isn't too large
(or if the quarks are sufficiently heavy), $q=1$ can be
a local minimum; {\it i.e.}, $q=1$ is {\it meta}stable.

The above carries
through for an arbitrary number of colors and flavors.
If metastable states arise, they are necessarily a $Z(N)$
state, with $\ell$ a (nontrivial) $N$th root of unity.
They are termed ``$Z(N)$ bubbles'' \cite{not,notnot}.

This all appears to be directly analogous to the usual problem
of metastable vacua, but there is one important difference.
For an ordinary potential, either in quantum mechanics or in field
theory, one never worries about the zero of the potential,
as that can be shifted at will.  
In the present case, however, the zero of the potential {\it is}
physical, and gives the free energy of the stable vacuum.
This is because the ``potential'' is multiplied by an overall
factor of $T^4$, and thermodynamically, derivatives with respect
to the temperature matter.

Thus there is no freedom to change the zero of the
potential for $q$.  For some $Z(N)$ bubbles, if
the potential at $q\neq 0$ is much higher than $q=0$,
it is well possible that the pressure in the bubble isn't
positive, but {\it negative}!
This was noticed first 
by Belyaev, Kogan, Semenoff, and Weiss \cite{not}.

This is a complete disaster.  I suggest a possible resolution.

When we deal with $q$, we are in fact dealing with an {\it angular}
variable.  If we write the potential for $q$ in terms of the
thermal Wilson line, it is
\beq
{\cal V}_{pert}(q) \; \sim \;
T^4 \; q^2\; (1-q)^2 \; \sim \; T^4 \; \tr \left(
(\log{\bf L})^2 \left( {\bf 1} \; - \; (\log{\bf L})^2 \right) \right) \; .
\eeq

This form is correct in perturbation theory, where at each
point in space, ${\bf L}(\vec{x})$ is an element of $SU(N)$.
As will become clear in the next section, however, if
we construct an effective theory for $\bf L$,
it no longer is an element of $SU(N)$.  Then
this potential, for the purely angular
part of $\bf L$, is ill defined when its modulus vanishes.
This ambiguity is easily cured by multiplying by
an overall factor of the modulus:
\beq
{\cal V}_{non-pert}({\bf L})  \; \sim \; T^4 \; \left( |\ell|^2
\; + \; \ldots \right) \tr 
\; \left(
(\log{\bf L})^2 \left( {\bf 1} \; - \; 
(\log{\bf L})^2 \right) \right) \; .
\label{potential1}
\eeq
This is rank conjecture: it is certainly
a {\it non}-perturbative modification of the potential.
There is no reason to exclude terms of higher order in 
$\sim |\ell|^2$.

This still does not solve the problem of the zero of the potential.
I now assume further that the 
Polyakov Loop Model (PLM) applies
\cite{loop1,loop2,loop3,loop4,loop5,loop6}.
Ignoring the angular variation in $\log {\bf L}$, the PLM potential is
\beq
{\cal V}_{PLM}(\ell)  \; \sim \; T^4 \left( 
b_2 \; |\ell|^2 \; + \; 
b_4 \; (|\ell|^2)^2 \right)  \; .
\label{potential2}
\eeq
The exact potential depends upon the
number of colors and flavors, {\it etc.}, but this is inessential here.
For $q=0$, the ``usual'' free energy
is given by minimizing ${\cal V}_{PLM}(\ell)$ with respect to $\ell$;
with the above convention, the ``mass''  squared for $\ell$ is negative
in the deconfined phase, $b_2 <0$, and positive in the confined
phase, $b_2 > 0$.

The complete potential is the sum of
${\cal V}_{non-pert}({\bf L})$ and ${\cal V}_{PLM}(\ell)$.
At fixed $\ell$, as before any metastable points are
elements of $Z(N)$.  The equation which determines $\ell$, however,
is changed, as any metastable state has an action which acts
like a {\it positive} mass term for $\ell$.  Thus the expectation
value of $\ell$ in a $Z(N)$ bubble  --- even deep in the deconfined
phase --- has $\ell < 1$, not $\ell =1$.
Before a $Z(N)$ bubble develops negative pressure, 
$\ell =0$, with zero pressure in the PLM.  More 
likely, $Z(N)$ bubbles become unstable in the $\ell$ direction
before $\ell = 0$.

This could be tested on the Lattice.  Compute in a theory in
which the splitting between the true vacuum and the metastable
state is small; (dynamical) heavy quarks will do.  
Then the expectation value of $\ell$ should
be smaller in the metastable state than in the stable vacuum.
This holds regardless of the question of renormalization
discussed in the next section.

If true, all perturbative calculations of
the lifetime of a metastable $Z(N)$ bubble are wrong
\cite{notnot}.  At best, they are a upper bound on the true
lifetime, which is somewhat useless.
On the other hand, it resurrects
the possibility that $Z(N)$ bubbles --- 
which spontaneously violate $CP$ symmetry --- might
have appeared in the early Universe.

\section{Renormalization of the Wilson Line}

The sections following this deal with mean field theory for the
thermal Wilson line.  Implicitly, this assumes that it is possible
to go from the bare Wilson line, as measured on the Lattice, to
the renormalized quantity.  How to do this on the Lattice is
presently an unsolved problem; in this section I review what
is known \cite{renorm_cont}.

In a pure gauge theory, the expectation value of a
closed Wilson loop, of length $L$ and area $A$, is
\beq
\langle \tr \; {\cal P} \;
\exp \left( i g \int A_\mu dx^\mu \right) 
\rangle \; = \; \exp\left(- m_0 L - \sigma A \right) \; . 
\eeq
The string tension, $\sigma$, is nonzero in the confined
phase, $T \leq T_c$, and vanishes in the deconfined phase, $T > T_c$.

The concern here is not with the term proportional to the area
of the loop, but with the length.  This is a type of mass renormalization
for an infinitely heavy quark.  For
example, it is easy to compute this to lowest order in perturbation
theory.  We are interested in an
ultraviolet divergent term, so over
short distances, it suffices to assume that the loop is straight.
For the sake of discussion, assume that the loop runs in the time
direction.  (New divergences arise when there are cusps in the loop; these
divergences can also be computed perturbatively, by a similar
analysis \cite{renorm_cont}.)  Then to lowest order, there is a contribution
\beq
\sim \; - g^2 \; \langle \int^\beta_0 \; d\tau_1 \;
\int^\beta_0 \; d\tau_2 \;
A_0(\vec{x},\tau_1) A_0(\vec{x},\tau_2) \rangle
\sim \; - \frac{g^2}{T} \; \int \; d^3 k \; \frac{1}{\vec{k}^2} \; .
\eeq
The integral is nominally divergent, but with either
dimensional or Pauli-Villars regularization, the divergence
vanishes \cite{renorm_cont}.  This cancellation is somewhat
trivial at one loop order, arising from having three powers of
momentum upstairs, and two powers downstairs.

Thus one would expect divergences to arise at $\sim g^4$, 
which are found.  However, for closed Wilson loops,
{\it all} such divergences
can be  absorbed into charge renormalization \cite{renorm_cont}.
This is a notable result:
in a quantum field theory, generally the renormalization of any
composite operator requires the calculation of its mixing with
all other operators with the same mass dimension and symmetries.
Like the action itself, however, the Wilson loop has a privileged
status; it doesn't mix with any other operator. 

As noted first by Polyakov \cite{renorm_cont},
this result can be understood on the basis of reparametrization
invariance for the Wilson loop.  We parametrize the loop as
a curve $x^\mu(s)$, where $s$ is the length along the path.
Then with $\dot{x}=dx^\mu/ds$, the term
\beq
\int \sqrt{\dot{x}^2}
\; ds 
\eeq
is invariant under $s \rightarrow s'(s)$.  Generally, physics
shouldn't depend upon how we label path length along the curve.
Because it has dimensions of length, however, the coefficient 
of this term must have dimensions of mass.  
With dimensional regularization, there is no such mass scale.
(The renormalization mass scale only enters to
ensure the proper running of the coupling constant.)
A term which has no mass dimension is
\beq
\int \sqrt{\ddot{x}^2}
\; ds \; ,
\eeq
$\ddot{x} = d^2x^\mu/ds^2$.
This is not reparametrization invariant, though,
and so does arise with dimensional regularization. 

On the other hand, assume that the regularization scheme {\it does}
introduce a mass scale.  On the Lattice, this is the inverse
lattice spacing, $\sim 1/a$.  Then a term proportional to the
length, $L$, does appear \cite{renorm_cont},
\beq
\sim \; - g^2 \; \frac{L}{a} .
\eeq
At nonzero temperature, $L/a=N_t$, 
the number of lattice steps in the time direction.
Clearly, this is the first term in an infinite series in the coupling
constant, $g$.

How to deal with the power divergences 
generated by the Lattice is at present an unsolved problem.
Since in the continuum there are neither logarithmic nor even
finite terms to worry about, this appears to be a technical,
albeit important, problem to solve.

Why is this important?
In the confined phase, this constant is not of any particular
consequence, as the Wilson loop is dominated by the string tension.
In the high temperature phase, however, the trace of the thermal
Wilson line is the order parameter for the phase transition.
It would be peculiar if a precise physical definition did not
exist.  
Any composite operator requires a condition to fix its renormalized
value; for the thermal Wilson line,
the natural prescription is that 
Polyakov loops approach one as $T\rightarrow \infty$.

To compute the leading perturbative correction to the
thermal Wilson line, it is necessary to include effects
from the Debye mass, $m_D \sim gT$.
Replacing the bare propagator for $A_0$, $1/\vec{k}^2$,  by 
$1/(\vec{k}^2 + m_D^2)$,
\beq
\sim \; - \frac{g^2}{T} \; \int \; d^3 k \; \frac{1}{\vec{k}^2 + m_D^2} \; .
\eeq
This divergent integral can be computed using either dimensional or
Pauli-Villars regularization.  Or, one can just subtract
the integral with $m_D=0$:
\beq
\; - \frac{g^2}{T} \; \int \; d^3 k \; 
\left( \frac{1}{\vec{k}^2 + m_D^2} - \frac{1}{\vec{k}^2}\right)
\; \sim - \frac{g^2}{T} \left( - m_D \right) \; \sim \;
+ \; g^3 \; .
\eeq
This was first demonstrated by  Gava and Jengo \cite{gava}.
That is, while the leading term is negative in the bare theory,
it is positive after regularization.
This change in sign is unremarkable, as the sign of a
renormalized operator is not preserved under regularization.  

This appears to indicate that the renormalized thermal Wilson
line is not a unimodular matrix.  One concern is that any quantity
$\sim g^3$ really arises from a two-loop graph, $\sim g^4$, times
an infrared singular piece $\sim 1/m_D \sim 1/g$.  Thus it is
necessary to ensure that the above is the only infrared singular
term at this order.  

A different calculation was performed by Korthals Altes \cite{altes}.
He computed the one loop corrections to the thermal Wilson line
in a background $A_0$ field.  The method is identical to that used
in sec. II to compute the interface tension.  
Classically, the thermal Wilson line is a special unitary matrix.
Korthals Altes finds
that the one-loop corrections to the thermal Wilson line 
are not only infinite (!), but generate a matrix which is neither
unitary nor special.  On the other hand, all Polyakov loops
are finite.

Thus even in the continuum, the renormalization of the thermal
Wilson line, and Polyakov loops, remains an unsolved problem.

A way of measuring renormalized Polyakov loops on the Lattice has been
proposed by Zantow {\it et al.} \cite{bielefeld_ren}.
They compute only two-point functions of the Polyakov
loop.  At short distances, the static potential can be
computed perturbatively, which allows one to extract the renormalized
Polyakov loop.  

\section{Deconfining Transition for Two, Three, and Four Colors}

\subsection{Polyakov Loops and Quarkless Baryons}

So far I have been concerned with the (pure glue)
theory at very high temperatures.  Now I turn to the
question of its behavior near the critical temperature.
I review Lattice results on the order of the
phase transition for two \cite{two}, three \cite{three}, and four 
\cite{four} colors, and then ask what constraints it places
on the mean field theory for Polyakov loops.  

Up to this point, I have only considered the trace of the thermal
Wilson line in the fundamental represenation, which is the the Polyakov loop
$\ell$.  By a local gauge
transformation, at each point in space one can  diagonalize the thermal
Wilson line.  These eigenvalues are gauge invariant, so since
${\bf L}(\vec{x})$ is an $SU(N)$ matrix, at each point
there are $N-1$ independent
degrees of freedom.  Another $N-1$ degrees of freedom are
given by the 
trace of powers of $\bf L$, $\tr {\bf L}^j$, $j=1\ldots(N-1)$.

Under a global $Z(N)$ transformation, the ``usual'' Polyakov loop
transforms as a field with charge-one, eq. (\ref{e222}); thus
I relabel it $\ell_1$.  Polyakov loops with 
higher $Z(N)$ charge are easy to construct.  
I introduce the traceless part of $\bf L$:
\beq
\widetilde{{\bf L}} = {\bf L} - \ell_1 {\bf 1} \; .
\eeq
Then I define the charge-two Polyakov loop  to be
\beq
\ell_2 = \frac{1}{N} \; {\rm tr} \, \widetilde{{\bf L}}^2 
\; = \; \frac{1}{N} \; {\rm tr} \, {\bf L}^2 
\; - \; \frac{1}{N^2} \; \left( {\rm tr} \, {\bf L} \right)^2
\; , 
\eeq
where
\beq
\ell_2 \rightarrow e^{2 i \phi} \ell_2 \; ,
\eeq
with $\phi$ as in eq. (\ref{phi}).
There are two operators with charge-two, $\ell_2$ and $\ell_1^2$.  

For example, consider two colors, and the parametrization of the
thermal Wilson line in the strict perturbative regime,
(\ref{e22a}).  The charge-one loop is
$\ell_1 = \cos(\pi q)$, eq. (\ref{e22b}), while the charge-two
loop is $\ell_2 = - \sin^2(\pi q)$.  At high temperature, where
$q =0,1$, $\langle\ell_1\rangle 
\rightarrow \pm 1$, while $\langle\ell_2\rangle \rightarrow 0$.

I note that Polyakov loops of charge-two and beyond are related
to the trace of the thermal Wilson line in higher $SU(N)$ 
representations.  For two colors, in perturbation theory
the trace of the
Wilson line in the adjoint representation is
$\tr ({\bf L}_{adj}) = 1 + 2 \ell_2$. So far, though, I haven't found this
particularly useful.

Continuing on, 
\beq
\ell_3 = \frac{1}{N} \; {\rm tr} \, \widetilde{{\bf L}}^3 \; 
\eeq
has charge three under the global $Z(N)$ symmetry.  
Other charge-three operators are $\ell_1^3$ and $\ell_1 \ell_2$.
The construction of operators with higher $Z(N)$ charge proceeds
similarly.  For example, operators with charge four, independent
of the singlet part, are given by 
${\rm tr} \, \widetilde{{\bf L}}^4$ and 
$({\rm tr} \, \widetilde{{\bf L}}^2)^2$, {\it etc.}

I stress that both the expectation values, and correlations
functions of, Polyakov loops of arbitrary charge are well
worth measuring on the Lattice.
When the $Z(N)$ symmetry is spontaneously broken at $T > T_c$,
Polyakov loops $\ell_j$ with charge $j=1\ldots(N-1)$
all acquire nonzero expectation values.
As Polyakov loops of charge-two and beyond are constructed from
the traceless part of the thermal Wilson line, they aren't that
interesting at high temperature; as $T \rightarrow \infty$, their expectation
values are proportional to nonzero powers of $g^2$, times powers of
$T$ to make up the mass dimension.  Near $T_c$, however,
there is nothing general which can be said about their expectation values.

I will assume that for a second order deconfining transition,
the only critical field is the charge-one loop \cite{svetitsky}.
Even so, Polyakov loops of charge-two and beyond will certainly
affect non-universal behavior.  One notable example is the 
Polyakov Loops Model
\cite{loop,loop1,loop2,loop3,loop4,loop5,loop6}, which
conjectures a relationship between the expectation value
of Polyakov loops and the pressure.  The
original model assumed that only the charge-one loop mattered,
but I no longer see why the expectation values of higher-charge
loops are not important as well.

There are certain Polyakov loops which have a privileged status:
these are those with charge-$N$.  As they are neutral under
$Z(N)$, their expectation values are nonzero at any
temperature.  I term such operators {\it quarkless baryons}.

In QCD, a baryon is $N$ quarks tied together through an antisymmetric
tensor in color space.  One can also consider a more general object,
a baryon ``junction'' \cite{baryon}.  This  is an antisymmetric color
tensor, with $N$ Wilson lines coming out of it.  Putting quarks
at the end of each line gives the usual QCD baryon, since
in the confined phase, Wilson lines are short, on the order of 
$\sim 1/\sqrt{\sigma}$, where $\sigma$ is the string tension.
While directly related to QCD baryons, 
without quarks, baryon junctions are not gauge invariant: only
junction anti-junction pairs are.

In contrast, all quarkless baryons are gauge invariant.  
The simplest quarkless baryon is $\ell_1^N$.  In mean field theory,
this is zero in the confined phase, and nonzero above.  In the full
quantum theory, $\langle \ell_1^N \rangle\neq 0$ at all $T$.
While there is not good Lattice data on this expectation value,
I assume that it is small below $T_c$, and large above, but this
is just a guess.  
Since junction anti-junction pairs involve $N$ Wilson lines, 
they are directly related to the operators for quarkless
baryons, $\ell_1^N$, {\it etc.}.  

\subsection{Effective Theories for Polyakov Loops}

I next turn to the construction of an effective theory
for the thermal Wilson line.  Remember how this proceeds
with an Ising model on a lattice.  
While the value of the spin on each site is $\pm 1$, 
after an effective spin is computed by averaging over
a domain of fixed size, the result effective spin is a continuous
variable, $\phi(\vec{x})$.  The effective theory is just the
usual $\phi^4$ theory.  By the renormalization group, it is
in the same universality class as the original Ising model.

The analogous proceedure can be carried through for the
thermal Wilson line.  
The effective thermal Wilson line, constructed by
a gauge invariant \cite{loop}
average over a domain of some size, is not an
$SU(N)$ matrix, but has more degrees of freedom.
I then consider {\it all} Polyakov loops, from
charge-one up to charge-$N$.
Of course there is no reason to stop there, but presumably
the number of effective fields which really matters is limited.

In an effective Lagrangian, the first thing to ask about 
are the mass terms:
\beq
\cle \; = \; m_1^2 |\ell_1|^2 \; + \; 
m_2^2 |\ell_2|^2 \; + \ldots
\eeq
The simplest assumption is that for $T \geq T_c$, condensation
is driven by a negative mass term for the charge-one loop:
\beq
m_1^2 < 0 \spc T \, > \, T_c \spc
m_1^2 > 0 \spc T \, < \, T_c \; ,
\eeq
and that the masses for all higher loops are positive at all
temperatures,
\beq
m_2^2 > 0 \spc m_3^2 > 0 \ldots 
\eeq
If so, then the charge-one loop controls the critical behavior 
\cite{svetitsky}.

There is good reason why one expects that
condensation is driven by that of the charge-one loop,
and {\it not} by loops with higher charge.
If the mass for the charge-one loop is
negative, the favored
vacuum is given by maximizing $|{\rm tr}{\bf L}|^2$.
After a global gauge rotation, we can always choose the
expectation value of $\bf L$ to be a diagonal matrix.
If ${\bf L}$ were a $U(N)$, instead of an $SU(N)$ matrix, 
then $|{\rm tr}{\bf L}|^2$ is maximized when $\bf L$ is
a constant phase times the unit matrix.  
This remains true when $\bf L$ is a $SU(N)$ matrix; for
it to be a unit matrix, however, it must be an element of
the center of the gauge group,
\beq
\langle {\bf L} \rangle \; = \; \ell_0 \;
\exp\left(i \phi\right) \; {\bf 1} \; .
\eeq
with $\phi$ a $Z(N)$ phase, $\phi = 2 \pi j/N$, $j=1\ldots (N-1)$.

For this particular expectation value, the vacuum does {\it not} spontaneously
break the (global) $SU(N)$ symmetry above $T_c$.  This accords
with naive expectation: the high temperature vacuum is not in a Higgs
phase.  The possibility of having an expectation value which doesn't
break $SU(N)$ is special to a field in the adjoint, as opposed
to the fundamental, representation.

On the other hand, assume that $m_1^2 >0$, and $m_2^2<0$ in the
deconfined phase, so that symmetry breaking
is driven by condensation of the charge-two,
instead of the charge-one, loop.  Then the vacuum is
given by maximizing $|{\rm tr}{\bf L}^2|^2$; this means that
the expectation value of $\bf L^2$ is an element of the center.
But if so, besides the $SU(N)$ invariant vacuum, there are also
vacua which are only invariant under $SU(N-1)$.  At present, there is no
evidence to suggest that the high temperature vacuum 
is one where $SU(N)$ spontaneously breaks to $SU(N-1)$.  

The above description can be extended beyond mean field theory,
at least if the deconfining
transition is of second order.  A transition driven by the charge-one
loop is one where $m_1^2 \rightarrow 0$ at $T_c$; one driven
by the charge-two loop is where $m_2^2\rightarrow 0$ at $T_c$, {\it etc.}
To be precise, for a transition driven by a charge-$k$ loop, both
its mass, and that of the charge-$(N-k)$ loop, vanish
at $T_c$.

The masses for Polyakov loops of all charges can be directly
measured on the Lattice.  
Even for the charge-one loop,
data near $T_c$ is, at present, limited \cite{gavai_masses,bielefeld_mass}.
There is also some data for higher charge loops for
three colors in $2+1$ dimensions \cite{twoplusone}.

Given this (crucial!) assumption about the masses, I next turn
to the order of the phase transition for a small number of colors.

\subsection{Two Colors: Second Order, and Quarkless Baryons}

For two colors, all Polyakov loops are real.
For the charge-one loop, I take the potential
\beq
\calv_1 \; = \; 
\frac{m_1^2}{2} \; \ell_1^2 \; + \; \frac{\lambda_1}{4} \; \ell_1^4\; ,
\eeq
with a positive quartic coupling, $\lambda_1 > 0$.  Of course higher
powers in $\ell_1$ are also possible.  Near a second order phase
transition, however, the most relevant operators, with the fewest powers
of $\ell_1$, dominate.

Invariant terms involving the charge-two loop include
\beq
{\cal V}_2 \; = \; h \; \ell_2 \; + \; \frac{1}{2} 
\; m_2^2 \; \ell_2^2\;+\;
\ldots
\eeq
plus terms $\sim \ell_2^3$, $\sim \ell_2^4$, {\it etc.}
All powers of $\ell_2$ are allowed because it is a singlet.
The potential which mixes the charge-one and charge-two loops starts
out as
\beq
{\cal V}_{mix} \; = \; \xi \; \ell_1^2 \; \ell_2  + \ldots \; .
\eeq
plus many other terms; this has the lowest mass dimension.

There is extensive Lattice data on the nature of the deconfining
phase transition \cite{two}.  Especially from the work of
Engels {\it et al}, it appears that the transition is of
second order.  To wit, the critical exponents 
are within $\sim 1 \%$ of the values expected for the Ising model
\cite{svetitsky}.

There is a surprise, however.  As first stressed by Damgaard \cite{two},
the expectation value of the Polyakov loop in the adjoint representation
is an approximate order parameter.  
In perturbation theory, the adjoint Polyakov loop
is $1+ 2 \ell_2$, so from the Lattice data,
the expectation value of $\ell_2$ presumably jumps at $T_c$.  

From the terms above, the expectation value of the charge-two loop is
\beq
\langle \ell_2 \rangle \; = \; - \; \frac{h + \xi^2 \, 
\ell_0^2}{m_2^2} \; .
\eeq
To explain the jump in $\langle \ell_2 \rangle$ about $T_c$,
there are then two possibilities.  If the charge-two loop
is heavy, then the coupling constant of the charge-two loop
to the charge-one loop, $\xi$, must be large.  This means
that changes in the density of the charge-two loop is driven
by condensation of the charge-one loop.

The other
possibility is that $h$ and $\xi$ are not especially large,
but that the charge-two loop becomes light near $T_c$.  
The latter doesn't violate universality, as long as the charge-two
loop isn't massless at $T_c$.

There is no lattice data on $\langle \ell_1^2 \rangle$.
I presume that as suggested by mean field theory, $\ell_1^2$
quarkless baryons are rare below $T_c$, and common above.  This
is seperate from the changes in $\ell_2$.

\subsection{Three Colors}

\subsubsection{A ``Nearly'' Second Order Transition}

In an asymptotically free gauge theory, it is
natural to form the ratio of the true pressure to that of an ideal gas.
In principle, positivity of entropy does not require this ratio to
be less than one.  In practice, Lattice data with improved actions finds
that this ratio is less than one at all temperatures \cite{karsch}.

Numerical simulations find
that for three colors, the deconfining transition is
of first order \cite{three}, in agreement with general
arguments \cite{svetitsky}. 
For a first order transition, the pressure is continuous at $T_c$,
but the energy density jumps.  Thus consider
the ratio of the jump in the energy density to that of an
ideal gas.

This ratio is not bounded by one.  To illustrate this, consider
a bag model.  Above $T_c$, the pressure is that of an ideal
gas, minus a bag constant, $b$:
\beq
p_{bag} \; = \; c_0 T^4 \; - b \; .
\eeq
The pressure is assumed to vanish below $T_c$, with 
$T_c$ fixed by $p_{bag} = 0$.  
This bag model does not describe the Lattice data near $T_c$,
but is a useful construct.
In the bag model, the ratio of energies is:
\beq
\frac{\delta e}{e_{ideal}}|_{Bag} \; = \; \frac{4}{3} \; .
\eeq
In contrast, Lattice data appears to give a result which is much smaller:
\beq
\frac{\delta e}{e_{ideal}}|_{Lattice} \; \sim \; \frac{1}{3} \; .
\eeq

The mass of the charge-one loop has also
been measured on the Lattice \cite{bielefeld_mass}.  
It goes from $m_1/T \sim 2.5 $ at $\sim 2 T_c$, 
down to $m_1/T \sim .25 $ at $\sim T_c^+$.  This decrease
in the screening mass, apparently by a factor of ten,
strongly suggests the in fact the transition is even weaker
than the above comparison with the bag model suggests.
Instead of weakly first order, I prefer to call the
deconfining transition  {\it nearly second order}.

(It is necessary to measure correlation 
functions of Polyakov loops to see this decrease.  
Masses measured from other operators, such as plaquettes,
do not decrease dramatically about $T_c$ \cite{gavai_masses}.  
This implies that the mixing between Polyakov loops and
and plaquette operators are small.  This small mixing is found
in related problems \cite{laineph}.)

An effective theory cannot explain why the deconfining transition
is weakly first order; it merely requires that certain coupling
constants are small.

\subsubsection{Polyakov Loops with Charge One and Minus One}

For three colors, I consider Polyakov loops with charge one, two and three.

There is no data on the expectation values for the quarkless
baryons, $\ell_1^3$, $\ell_1 \ell_2$, and $\ell_3$.  I presume
that as indicated by mean
field theory, $\langle \ell_1^3 \rangle$ is small below
$T_c$, and large above.  It would be interesting to knowhow
the density of the other quarkless
baryons, $\ell_1 \ell_2$ and $\ell_3$, change about $T_c$.
Regardless of the renormalization issues discussed in sec. (4),
changes in these expectation values are presumably physical.

Thus I concentrate on the interaction between the charge-one
and the charge-two loops.  Remember that for three (or more) colors,
the $\ell_j$'s are all complex valued fields.  
The potential for charge-one loops
is dictated by the global $Z(3)$ symmetry:
\beq
{\cal V}_1 \; = \; m_1^2 \; |\ell_1|^2 \; + \;
\kappa_1\;  \left( \ell_1^3 \; + \; (\ell_1^*)^3 \right) \; + \;
\lambda_1 \; \left( |\ell_1|^2 \right)^2 \; .
\eeq
The notable feature is the appearance of a cubic term, which
necessarily ensures a first order transition \cite{svetitsky}.

The charge-two loop has charge minus one under $Z(3)$, so
its potential is the same, albeit with different masses and
coupling constants:
\beq
{\cal V}_2 \; = \; m_2^2 \; |\ell_2|^2 \; + \;
\kappa_2\;  \left( \ell_2^3 \; + \; (\ell_2^*)^3 \right) \; + \;
\lambda_2 \; \left( |\ell_2|^2 \right)^2 \; .
\eeq

There are many terms by which the charge-one and charge-two fields
can mix.  The most important is that with the smallest mass dimension:
\beq
{\cal V}_{mix} \; = \; \xi \left( \ell_1 \ell_2 \; + \;
\ell_1^* \ell_2^* \right) \; .
\eeq
In terms of the original thermal Wilson line, this term
is $\sim (\tr {\bf L} )(\tr {\bf L}^2)$, {\it etc.}

If the charge-two field is heavy 
near $T_c$, we can integrate it out.
While it may be a mess to do so analytically,
any resulting potential, involving only $\ell_1$, must still respect
the overall $Z(3)$ symmetry.  This produces
a potential identical
in form to ${\cal V}_1$, but with different values for the mass
and coupling constants.  A weakly first order requires that
the effective cubic coupling in the resulting effective theory
is small, $\widetilde{\kappa}_1\ll 1$.  

If the charge-two loop becomes light near $T_c$,
and if it mixes strongly with the charge-one loop through
a large coupling constant $\xi$, then its effects cannot
be neglected.  

There is another possibility.  The mass and quartic terms in
the potentials are invariant not just under $Z(3)$, but under
a global $U(1)$ symmetry.  Assume that the charge-two field
is always heavy.  Then all terms in both potentials are invariant
under a global $U(1)$, with the exception of the cubic terms,
with couplings $\kappa_1$ and $\kappa_2$, {\it and} the mixing
term between $\ell_1$ and $\ell_2$, with coupling $\xi$.  Thus
perhaps {\it all} terms invariant under $Z(3)$, but not $U(1)$,
are small.  That is, the heavy charge-two loop mixes weakly
with the charge-one loop.  This doesn't explain why all
$Z(3)$ couplings are small, but hints at a more general principle.

It will be interesting to see what detailed numerical studies
on the Lattice tell us.

\subsection{Four Colors: First Order from Charge-Two Loops}

For four colors, I consider just the charge-one and charge-two loops.
Under $Z(4)$, $\ell_1 \rightarrow i \ell_1$, so the 
potential for the charge-one field alone is
\beq
{\cal V}_1 \; = \; 
\frac{m_1^2}{2} \; |\ell_1|^2 \; + 
\; \lambda_1 \; (|\ell_1|^2)^2 \;
+ \; \kappa_1 \; \left(\ell_1^4 \; + \; (\ell_1^*)^4\right) \; .
\eeq
The term $\sim \lambda_1$ is $O(2)$ invariant,
while that $\sim \kappa_1$ is only invariant under $Z(4)$.

Under $Z(4)$, $\ell_2 \rightarrow - \ell_2$, so the potential
for the charge-two field by itself is just like that for the
charge-one field:
\beq
{\cal V}_2 \; = \; + \;
m_2^2 \; |\ell_2|^2 \; + 
\; \lambda_2 \; \left(|\ell_2|^2\right)^2 \;
+ \; \kappa_2 \; \left(\ell_2^4 \; + \; (\ell_2^*)^4 \right) \; .
\eeq
The allowed terms which mix the two fields are:
\beq
{\cal V}_{mix} \; = \; 
\zeta_1 \; \left( \ell_2^* \, \ell_1^2 \; + \; 
\ell_2 \, (\ell_1^*)^2 \right) \;
+ \; \zeta_2 \; \left( \ell_2 \, \ell_1^2 + \ell_2^* \, (\ell_1^*)^2 \right)
\; .
\eeq

Unlike two colors, a term linear in $\ell_2$ is not allowed by
the $Z(4)$ symmetry.  For $\zeta_2 = 0$, 
and assuming that the charge-two field
reamins heavy about $T_c$,
$\ell_2$ can be integrated out to give:
\beq
\; \sim \; 
- \; \frac{\zeta_1^2}{m_2^2} \; (|\ell_1|^2)^2 \; .
\eeq
One can convince oneself that this term is necessarily negative.
That is, the quartic coupling for the charge-one field
is shifted downward:
\beq
\widetilde{\lambda}_1
\; \equiv \; 
\lambda_1 - \frac{\zeta_1^2}{m_2^2} \; .
\eeq
The same holds if both $\zeta_1$ and $\zeta_2$ are nonzero:
integrating out the charge-two field generates corrections
to the quartic coupling constants of the charge-one field
which are uniformly negative, shifting them downwards.

The transition for four colors
appears to be of first order \cite{four}.  One explanation
for this is that the original coupling constants
$\lambda_1$ and $\lambda_2$ are positive, but after
integrating out the charge-two field, they become negative,
and thus drive the transition first order.

The analysis for higher numbers of colors is then immediate.
The leading term which couples a charge-$j$ loop to the charge-one
loop is
\beq
{\cal V}_{mix} \; = \; 
\xi \left( \ell_j^* \, \ell_1^j \; + \; 
\ell_j \, (\ell_1^*)^j \right) \; .
\eeq
Assuming that the charge-$j$ loop is heavy about $T_c$, we can
integrate it out, which produces a term in the potential
for the charge-one loop $\sim (|\ell_1|^2)^j$.  For charges
greater than two, this is less relevant (has smaller mass dimension)
than the quartic terms expected to dominate.  

Thus if the deconfining transition is of first order for more than
four colors, in the present language
it is uniquely due to how the coupling between charge-two and
charge-one loops affects the effective quartic coupling constant
for the charge-one loop.  Polyakov loops with charge greater than
two do not affect the order of the transition.

\section{A Parting Comment}

While it is true that, in equilibrium,
all thermodynamic quantities follow from the pressure,
experience with the perturbative calculation of processes near
equilibrium ---
such as transport coefficients,  real photon production,
{\it etc.} --- teaches us they often depend
on the details of equilibrium correlation functions.
Thus regardless of theoretical prejudice,
such as the Polyakov Loops Model, it is important to measure as many
gauge invariant correlation functions as possible.  

There is an astounding amount of superb data which is pouring out of
RHIC \cite{exp}.  Many features, including the change in the spectrum of
``hard'' particles, details of Hanbury-Brown-Twiss interferometry,
chemical composition, {\it etc.}, appear to defy explanation 
by any conventional mechanisms \cite{miklos}.  We may very well need a 
detailed understanding of
the theory, near $T_c$ and above, in order to sort out these amazing results.

\end{document}